\documentclass[a4paper,english,aps,manuscript,aps,preprint,nofootinbib]{revtex4}
\usepackage[T1]{fontenc}
\usepackage[latin9]{inputenc}
\usepackage{color}
\usepackage{babel}
\usepackage{amsmath}
\usepackage{amssymb}
\usepackage{graphicx}
\usepackage[unicode=true,
 bookmarks=false,
 breaklinks=false,pdfborder={0 0 1},backref=false,colorlinks=true]
 {hyperref}

\makeatletter

\pdfpageheight\paperheight
\pdfpagewidth\paperwidth

\@ifundefined{textcolor}{}
{%
 \definecolor{BLACK}{gray}{0}
 \definecolor{WHITE}{gray}{1}
 \definecolor{RED}{rgb}{1,0,0}
 \definecolor{GREEN}{rgb}{0,1,0}
 \definecolor{BLUE}{rgb}{0,0,1}
 \definecolor{CYAN}{cmyk}{1,0,0,0}
 \definecolor{MAGENTA}{cmyk}{0,1,0,0}
 \definecolor{YELLOW}{cmyk}{0,0,1,0}
}

\makeatother

\begin{document}
\preprint{FTUV-19-1219.6959, IFIC/19-61}
\title{Composite Higgs bosons from neutrino condensates\\
in an inverted see-saw scenario}
\author{Leonardo Coito }
\email{leonardo.coito@uv.es}

\affiliation{Departament de Física Teòrica, Universitat de València\\
 and IFIC, Universitat de València-CSIC\\
Dr. Moliner 50, E-46100 Burjassot (València), Spain}
\author{Carlos Faubel }
\email{carlos.faubel@uv.es}

\affiliation{Departament de Física Teòrica, Universitat de València\\
 and IFIC, Universitat de València-CSIC\\
Dr. Moliner 50, E-46100 Burjassot (València), Spain}
\author{Arcadi Santamaria}
\email{arcadi.santamaria@uv.es }

\affiliation{Departament de Física Teòrica, Universitat de València\\
 and IFIC, Universitat de València-CSIC\\
Dr. Moliner 50, E-46100 Burjassot (València), Spain}
\begin{abstract}
We present a realization of the idea that the Higgs boson is mainly
a bound state of neutrinos induced by strong four-fermion interactions.
The conflicts of this idea with the measured values of the top quark
and Higgs boson masses are overcome by introducing, in addition to
the right-handed neutrino, a new fermion singlet, which, at low energies,
implements the inverse see-saw mechanism. The singlet fermions also
develop a scalar bound state which mixes with the Higgs boson. This
allows us to obtain a small Higgs boson mass even if the couplings
are large, as required in composite scalar scenarios. The model gives
the correct masses for the top quark and Higgs boson for compositeness
scales below the Planck scale and masses of the new particles above
the electroweak scale, so that we obtain naturally a low-scale see-saw
scenario for neutrino masses. The theory contains additional scalar
particles coupled to the neutral fermions, which could be tested in
present and near future experiments. 

\end{abstract}
\maketitle

\section{Introduction}

In 1989 Bardeen, Hill and Lindner \citep{Bardeen:1989ds} (BHL) put
forward the idea that the Higgs boson could be a bound state of top
quarks by using an adapted Nambu \& Jona-Lasinio (NJL) model \citep{Nambu:1961fr,Nambu:1961tp}
(see also \citep{Miransky:1988xi,Miransky:1989ds,Suzuki:1989nv,Suzuki:1989si,Marciano:1989mj,Marciano:1989xd}
for related approaches). The mechanism is very attractive because
it gives a prediction for the top quark mass and for the Higgs boson
mass, which can be compared with experiment. These predictions are
based on two main ingredients: i) The existence of a Landau pole in
the top quark Yukawa and Higgs boson self-couplings at the compositeness
scale. ii) The existence of infrared fixed points in the renormalization
group equations (RGE), which make the low energy predictions stable
\citep{Hill:1980sq,Hill:1985tg}. Unfortunately, the minimal version
predicts a too heavy top quark (mass above $200$ GeV) and an extremely
heavy Higgs boson ($m_{h}\sim2m_{t}$ at leading order, and above
$300$ GeV once corrections are included). Since then, many authors
tried to generalize the mechanism to give predictions in agreement
with experiment (for a review see for instance \citep{Cvetic:1997eb,Hill:2002ap}).

Among the different ideas we find particularly interesting the possibility
that the Higgs boson is, mainly, a bound state of neutrinos \citep{Krog:2015cna,Smetana:2013hm,Antusch:2002xh,Martin:1991xw}
because, after all, neutrinos are already present in the Standard
Model (SM) and should have some non-SM interactions in order to explain
the observed neutrino masses and mixings. In particular, if neutrino
masses come from the type I see-saw model, neutrino Yukawa couplings
could be large enough to implement the BHL mechanism. This approach
has two important problems: i) In the type I see-saw the Majorana
masses of right-handed neutrinos should be quite large (at least $\sim10^{13}$
GeV) for Yukawa couplings of order one, which are needed to generate
the bound state. This means that there are just a few orders of magnitude
of running to reach the Landau pole before the Plank scale. ii) It
is very difficult to obtain the $\text{125}$~GeV Higgs boson mass
because it tends to be too heavy. In Ref.~\citep{Krog:2015cna} problem
i) was circumvented by adding three families of neutrinos with identical
couplings and problem ii) by adding, by hand, a fundamental scalar
singlet which mixes with the Higgs doublet. This produces a shift
in the Higgs boson mass and allows one to accommodate the measured
value.

Here, we propose a quite different approach: i) will be solved by
lowering the right-handed neutrino mass. This can be implemented naturally
in inverse see-saw\footnote{There is some recent work in which low-scale neutrino see-saw models
are embedded in the composite scalar scenario, see \citep{Dib:2019jod}.
However, in this work the Higgs boson doublet is a fundamental Higgs
and no attempt is made to explain the observed Higgs boson and top
quark masses. Rather, the NJL framework is used to justify lepton
number violation and provide solutions to the cosmological baryon
asymmetry and dark matter problems (for the use of right-handed neutrino
condensates to solve these problems see also \citep{Barenboim:2016xhn,Barenboim:2010nm,Barenboim:2008ds}).} type scenarios \citep{Mohapatra:1986aw,Mohapatra:1986bd} (see also
\citep{Wyler:1982dd} and \citep{Bernabeu:1987gr}). To solve ii)
we will also introduce a new scalar, however, this scalar will be
a composite of the new fermions required in the inverse see-saw scenario,
therefore, its couplings will be fixed by the compositeness condition. 

Thus, in Sec.~\ref{sec:The-BHL-mechanism} we briefly describe the
BHL mechanism: in Subsec.~\ref{subsec:The-SM-case} we sketch the
minimal version as applied to the pure SM and in Subsec.~\ref{subsec:The-Krog-Hill}
we present the case in which the Higgs boson is mainly a bound state
of neutrinos within the type I see-saw scenario according to Ref.~\citep{Krog:2015cna}.
In Sec.~\ref{sec:the-model} we discuss our implementation of the
BHL mechanism. First, in Subsec.~\ref{subsec:The-inverse-see-saw},
we briefly review the inverse see-saw model. Then, in Subsec.~\ref{subsec:njl-iss-description}
we give the high energy Lagrangian, which only contains fermions and
three four-fermion interactions, and derive the low energy Lagrangian,
which contains the SM Higgs doublet as a bound state of the fermions
plus and additional composite scalar singlet. We also obtain the matching
conditions for the couplings of the two Lagrangians and show that
all the dimensionless couplings of the low energy Lagrangian (three
Yukawa and three quartic couplings) are written in terms of two parameters
at the compositeness scale. Last, in Subsec.~\ref{subsec:top-higgs-masses}
we run all couplings down to the electroweak scale and compute the
top quark and Higgs boson masses, which are compared with the experimental
values. Finally, in Sec.~\ref{sec:Conclusions} we discuss the main
results of this work.

\section{The BHL mechanism\label{sec:The-BHL-mechanism}}

\subsection{The SM case\label{subsec:The-SM-case}}

In the BHL approach one considers a SM without the scalar doublet
and, instead, one introduces a four-fermion interaction among top
quarks

\begin{equation}
\mathcal{L}_{4f}=\frac{h_{t}^{2}}{m_{0H}^{2}}\left(\overline{T}_{L}t_{R}\right)\left(\overline{t}_{R}T_{L}\right)+\,\mathrm{h.c.}\,,\label{eq:SM4fLagrangian}
\end{equation}
where $T_{L}$ is the SM left-handed third generation quark doublet
and $t_{R}$ is the top quark right-handed singlet. By iterating this
interaction one can show \citep{Nambu:1961tp,Bardeen:1989ds} that
if it is strong enough it will induce SSB, $\langle\overline{t}_{L}t_{R}\rangle\not=0$
and the presence of a scalar bound state of top quarks. For our purposes,
this can be seen more transparently by using the ``bosonized'' version
\citep{Bardeen:1989ds}, namely, the four-fermion interaction can
be written as
\begin{equation}
\mathcal{L}_{\Lambda}=-m_{0H}^{2}|H|^{2}+y_{0t}\overline{T}_{L}t_{R}H+\mathrm{h.c.}\,,\label{eq:SMLambdaLagrangian}
\end{equation}
where $H$ is a scalar doublet. On can easily check the equivalence
of Eq.~\eqref{eq:SM4fLagrangian} and Eq.~\eqref{eq:SMLambdaLagrangian}
by using the equations of motion to remove the scalar field $H$,
which gives $h_{t}^{2}=y_{0t}^{2}$.

This equivalence is exact, at some scale $\Lambda$, because the field
$H$ has no kinetic term and, at this point, it must be seen as an
auxiliary field. However, quantum corrections involving only fermion
loops will necessarily generate a scalar kinetic term and scalar self-interactions.
Thus, at a scale $\mu$ just below the $\Lambda$ scale one generates
kinetic terms for the $H$, a renormalization of the mass and quartic
terms 
\begin{equation}
\mathcal{L}_{\mu}=Z_{H}(\mu)|D_{\mu}H|^{2}-\tilde{m}_{H}^{2}(\mu)|H|^{2}-\frac{1}{2}\tilde{\lambda}(\mu)|H|^{4}+\tilde{y}_{t}(\mu)\overline{T}_{L}t_{R}H+\mathrm{h.c.}\label{eq:SMmuLagrangian}
\end{equation}
Calculation of the corresponding fermion loops with a cutoff $\Lambda$
and imposing the ``compositeness'' boundary conditions
\begin{equation}
Z_{H}(\Lambda)=0\,,\quad\tilde{\lambda}(\Lambda)=0\,,\quad\tilde{m}_{H}^{2}(\Lambda)=m_{0H}^{2}\:,\quad\tilde{y}_{t}(\Lambda)=y_{0t}\,,\label{eq:SMboundaries-aux}
\end{equation}
one obtains
\begin{equation}
Z_{H}(\mu)=N_{c}y_{0t}^{2}L(\mu)\,,\quad\tilde{\lambda}(\mu)=2N_{c}y_{0t}^{4}L(\mu)\,,\quad\tilde{m}_{H}^{2}(\mu)=m_{0H}^{2}-2y_{0t}^{2}\frac{N_{c}}{16\pi^{2}}\left(\Lambda^{2}-\mu^{2}\right)\,,\label{eq:SMmatching}
\end{equation}
where
\begin{equation}
L(\mu)\equiv\frac{1}{16\pi^{2}}\log\frac{\Lambda^{2}}{\mu^{2}}\,.\label{eq:Log-def}
\end{equation}

Notice that Yukawa couplings $\tilde{y}_{t}$ do not receive one-loop
corrections from fermions, therefore $\tilde{y}_{t}(\mu)=\tilde{y}_{t}(\Lambda)=y_{0t}$.

Then, one re-scales the field $H\rightarrow H/\sqrt{Z_{H}(\mu)}$
to obtain the SM Lagrangian (with only top quark Yukawa couplings)
\begin{equation}
\mathcal{L}_{\mu R}=|D_{\mu}H|^{2}-m_{H}^{2}(\mu)|H|^{2}-\frac{1}{2}\lambda(\mu)|H|^{4}+y_{t}(\mu)\overline{T}_{L}t_{R}H+\mathrm{h.c.\,,}\label{eq:SMRmuLagrangian}
\end{equation}
with 
\begin{equation}
m_{H}^{2}(\mu)=\tilde{m}_{H}^{2}(\mu)/Z_{H}(\mu)\:,\qquad y_{t}^{2}(\mu)=y_{0t}^{2}/Z_{H}(\mu)=\frac{1}{N_{c}L(\mu)}\,,\label{eq:SMpars1}
\end{equation}
\begin{equation}
\lambda(\mu)=\tilde{\lambda}(\mu)/Z_{H}^{2}(\mu)=\frac{2}{N_{c}L(\mu)}=2y_{t}^{2}(\mu)\,.\label{eq:SMpars2}
\end{equation}
We see that the two couplings, $y_{t}(\mu)$ and $\lambda(\mu)$,
diverge together when $\mu=\Lambda$, $\lambda(\mu)=2y_{t}^{2}(\mu$).
The last equation is very important since it gives the relation between
the Higgs boson and the top quark masses. In fact, if $H^{0}=(v+h)/\sqrt{2}$
one finds
\begin{equation}
m_{t}=y_{t}(m_{t})\frac{v}{\sqrt{2}}\:,\quad m_{h}^{2}=\lambda(m_{t})v^{2}=2y_{t}^{2}(m_{t})v^{2}=4m_{t}^{2}\,,\label{eq:SMmh2mt}
\end{equation}
which is the standard compositeness result. Moreover, once $\Lambda$
is given we have a prediction for the top quark and Higgs boson masses
(for simplicity we take $m_{h}\approx\lambda(m_{t})v^{2}$, the small
running from $m_{t}$ to $m_{h}$ and finite corrections can also
be included).
\begin{equation}
y_{t}^{2}(m_{t})=\frac{1}{N_{c}L(m_{t})}=\frac{16\pi^{2}}{N_{c}\log(\Lambda^{2}/m_{t}^{2})}\;\rightarrow\quad m_{t}^{2}=\frac{8\pi^{2}v^{2}}{N_{c}\log(\Lambda^{2}/m_{t}^{2})}\,.\label{eq:SMmt}
\end{equation}
The solution can be written in terms of the Lambert function $W_{-1}(x)$
\begin{equation}
m_{t}=\Lambda\exp\left(\frac{1}{2}W_{-1}\left(-\frac{8\pi^{2}v^{2}}{N_{c}\Lambda^{2}}\right)\,\right)\label{eq:mt-largeNC}
\end{equation}
and gives $m_{t}=164$ GeV for $\Lambda=10^{15}$ GeV, which is reasonable,
while the prediction $m_{h}\text{\ensuremath{\sim2m_{t}}}$ is quite
wrong. For lower $\Lambda$, $m_{t}$ (and so $m_{h}$) is larger.
For instance, if $\Lambda=10^{10}$~GeV, Eq.~\eqref{eq:mt-largeNC}
gives $m_{t}=210$~GeV. However, this calculation is no complete.
Eqs.~\eqref{eq:SMpars1} should be understood as boundary conditions
for scales close to the compositeness scale $\Lambda$, where the
Higgs boson is not a dynamical field (therefore, cannot appear in
loops), and gauge corrections are presumably small (the main contributions
come from QCD, which are small at large scales). Thus, below the compositeness
scale the Higgs boson contributions (fermion self-energies, vertex
corrections and scalar self-interactions) should be included. Also
strong interactions could become important at lower energies. Therefore,
to give accurate predictions one should use the complete RGE of the
SM with the boundary conditions Eqs.~(\ref{eq:SMpars1}--\ref{eq:SMpars2}).
Still the calculation above illustrates the main consequences of the
approach; once $\Lambda$ is given everything is fixed, in particular
$m_{t}$ and $m_{h}$.

Let us now see how the full predictions can be obtained. The complete
SM RGE beta functions are

\begin{align}
\beta_{y_{t}} & =y_{t}\left(\frac{9}{2}y_{t}^{2}-8g_{3}^{2}-\frac{9}{4}g_{2}^{2}-\frac{17}{20}g_{1}^{2}\right)\,,\label{eq:RGESM-yt}\\
\beta_{\lambda} & =12\left(\lambda^{2}+\frac{9}{400}g_{1}^{4}+\frac{3}{40}g_{1}^{2}g_{2}^{2}+\frac{3}{16}g_{2}^{4}+\lambda\left(y_{t}^{2}-\frac{3}{20}g_{1}^{2}-\frac{3}{4}g_{2}^{2}\right)-y_{t}^{4}\right)\,,\label{eq:RGESM-lambdaH}
\end{align}
where $g_{3}$, $g_{2}$, $g_{1}$ are the SM SU(3), SU(2), U(1) SM
gauge couplings (normalized with the SU(5) prescription, such that
the weak mixing angle is given by $\tan^{2}\theta_{W}=(3/5)g_{1}^{2}/g_{2}^{2}$
and, for a generic coupling $g$, we are using the convention $\beta_{g}=16\pi^{2}\mu dg/d\mu$).

One can check that the couplings in Eq.~\eqref{eq:SMpars1} satisfy
these equations once one takes $N_{c}=3$, removes gauge terms and
includes only contributions from fermion loops.

To impose the boundary conditions in Eq.~\eqref{eq:SMpars1} we cannot
take directly $\mu=\text{\ensuremath{\Lambda}}$ since then the couplings
diverge. We will take the boundary conditions slightly below $\Lambda$,
at $\mu=\Lambda_{\kappa}\equiv\Lambda/\kappa$ with $\kappa\gtrsim1$,
\begin{equation}
y_{t}^{2}(\Lambda_{\kappa})=\frac{8\pi^{2}}{N_{c}\log(\kappa)}\,,\quad\lambda(\Lambda_{\kappa})=2y_{t}^{2}(\Lambda/\kappa)=\frac{16\pi^{2}}{N_{c}\log(\kappa)}\,,\label{eq:SMboundaries}
\end{equation}
which can be seen as matching conditions between the SM and the theory
at the compositeness scale. Thus, we are assuming that we have the
complete SM below $\Lambda_{\kappa}\lesssim\Lambda$, while from $\Lambda_{\kappa}$
to $\Lambda$ we have an effective theory in which the Higgs boson
is not a dynamical degree of freedom (does not run in loops) and gauge
interactions are neglected. This setup eventually leads to a Landau
pole for all couplings at the scale $\Lambda$, but introduces a dependence
on the parameter $\kappa$ which parametrizes possible matching uncertainties
at the scale $\Lambda$. Most of these uncertainties will be erased
in the running from $\Lambda$ to the electroweak scale, if $\Lambda\gg m_{t}$,
because of the infrared fixed point structure of Eqs.~(\ref{eq:RGESM-yt}--\ref{eq:RGESM-lambdaH}).
Notice that the details of the complete theory (in this case the use
of four-fermion interactions to obtain the bound states) are encapsulated
in the boundary conditions Eq.~\eqref{eq:SMboundaries}.

Thus, one takes the gauge couplings measured at the $Z$-boson mass
scale, $m_{Z}$, runs them up to the scale $\Lambda_{k}$ and then,
using Eq.~\eqref{eq:SMboundaries} and Eqs.~(\ref{eq:RGESM-yt}--\ref{eq:RGESM-lambdaH}),
one obtains $y_{t}(m_{t})$ and $\lambda(m_{h})$, and therefore\footnote{Well known SM finite corrections at the weak scale can also be included,
if necessary.}, $m_{t}$ and $m_{h}$.

With this procedure one gets $m_{t}=223\pm3$~GeV, $m_{h}=246\pm4\,$GeV
for $\Lambda=10^{17}$ GeV and $m_{t}=455\pm45$~GeV, $m_{h}=605\pm142\,$GeV
for $\Lambda=10^{4}$ GeV, where the uncertainties come from the input
parameters (basically $g_{3}(m_{Z})$) and $\kappa$, which we vary
$\kappa$ from $2$ to $10$. These values are compatible with the
results of Ref.~\citep{Bardeen:1989ds} when their input parameters
are used. This has to be compared with the measured values, $m_{t}\sim173$~GeV
and $m_{h}\sim125$~GeV. Quarks, however, are not observed as free
particles and there are several possible definitions for their masses.
The most precisely measured value for the top quark mass is obtained
by kinematic reconstruction and yields $m_{t}\sim173$~GeV. Its connection
with the parameters of the Lagrangian is not clear, although it is
believed to be related, up to corrections of the order of the QCD
scale, $\Lambda_{\mathrm{QCD}}$, with the so called pole mass, denoted
here by $m_{t}$ and defined as the position of the pole of the propagator
computed perturbatively. Running Yukawa couplings are defined in the
$\overline{\mathrm{MS}}$ scheme and are more closely related to the
running mass $\overline{m}_{t}(\overline{m}_{t})$. It is known that
the relation between these two quantities, pole and running masses,
is affected by large QCD radiative corrections which produces a shift
of the order of $10$~GeV between the two definitions (see, for instance
Ref.~\citep{Kniehl:2014yia}). This would lower the mass from $173$
GeV to $163\,$GeV. The situation is even more complicated if one
also includes electroweak corrections, which can be large because
of the presence of tadpole contributions \citep{Kniehl:2014yia}).
Fortunately one can show that, at least at one loop, the connection
between the pole mass and the Yukawa coupling is free from these tadpole
contributions, rendering the electroweak corrections small. Therefore,
we use the known expressions that connect the pole quark mass with
the Yukawa coupling \citep{Kniehl:2014yia,Hempfling:1994ar}). Anyway,
even taking into account all these corrections it is clear that the
top quark mass prediction is off by more than $50$ GeV. The Higgs
mass prediction is even worse since the measured value is $m_{h}=125.1\pm0.14$~GeV
and its connection to quartic couplings is only affected by small
weak corrections.

Clearly the minimal version of the mechanism is off even for scales
close to the Planck scale. The Higgs boson mass prediction, above
the top quark mass, seems particularly difficult to reconcile with
experiment.

\subsection{The Higgs as a neutrino bound state\label{subsec:The-Krog-Hill}}

Here we briefly review the scenario in which the Higgs boson is a
bound state of neutrinos \citep{Krog:2015cna,Smetana:2013hm,Antusch:2002xh,Martin:1991xw},
specifically, we follow more closely the Krog\&Hill (KH) approach
\citep{Krog:2015cna}. KH introduce the following four-fermion interactions
(they assume 3 families of leptons with a common coupling and 3 colours
for the quarks, whose indices will not be displayed explicitly)

\begin{equation}
\mathcal{L}_{4f}=\frac{h_{\nu}^{2}}{m_{0H}^{2}}\left(\overline{L}_{L}\nu_{R}\right)\left(\overline{\nu}_{R}L_{L}\right)+\frac{h_{t\nu}^{2}}{m_{0H}^{2}}\left(\overline{L}_{L}\nu_{R}\right)\left(\overline{t}_{R}T_{L}\right)+\frac{1}{2}\overline{\nu_{R}^{c}}M_{R}\nu_{R}\mathrm{\,+\,h.c.},\label{eq:KH4fLagrangian}
\end{equation}
where $L_{L}$ are the left-handed lepton doublets, $\nu_{R}$ are
the right-handed neutrinos and $M_{R}$ is a $3\times3$ right-handed
neutrino Majorana mass matrix necessary to implement the type I see-saw
mechanism for neutrino masses. If $h_{\nu}\gg h_{\nu t}$, the NJL
interaction can be written in terms of an auxiliary scalar doublet
$H$ (to be identified as the Higgs doublet)

\begin{equation}
\mathcal{L}_{\Lambda}=-m_{0H}^{2}H^{\dagger}H+y_{0t}\overline{T}_{L}t_{R}H+y_{0\nu}\overline{L}_{L}\nu_{R}H+\frac{1}{2}\overline{\nu_{R}^{c}}M_{R}\nu_{R}+\mathrm{h.c.}\,,\label{eq:KHLagrangianLambda}
\end{equation}
as can be checked by removing the field $H$ using the equations of
motion. Notice that in this procedure one neglects terms of order
$y_{0t}^{2}$ which would induce a pure top quark four-fermion interaction.
This means that $H$ will be mainly a bound state of neutrinos with
a small contribution from top quarks.

Below the scale $\Lambda$ a Higgs kinetic term and a potential are
generated and, after Higgs wave function renormalization, one recovers
a SM Lagrangian, Eq.~\eqref{eq:SMRmuLagrangian}, but including neutrino
Yukawa couplings and right-handed neutrino Majorana masses, written
in terms of renormalized couplings $y_{t}(\mu)$ and $y_{\nu}(\mu)$.
Then, one runs this Lagrangian from $\Lambda$ to the scale $M_{R}$,
where right-handed neutrinos decouple and generate active neutrino
masses given by the standard see-saw formula $m_{\nu}\sim y_{\nu}^{2}(M_{R})\left\langle H\right\rangle /M_{R}$.
Since $y_{\nu}(M_{R})$ are expected to be $\mathcal{O}(1)$ in order
to drive the NJL and $m_{\nu}<1$~eV, $M_{R}$ is expected to be
larger than $10^{13}$~GeV. Below $M_{R}$ one has the SM with tiny
neutrino Majorana masses. The point is that, above the scale $M_{R}$
the neutrino Yukawa coupling, $y_{\nu}$, contributes to the running
of the top quark Yukawa coupling
\begin{equation}
\beta_{y_{t}}=y_{t}\left(\frac{9}{2}y_{t}^{2}+3y_{\nu}^{2}+\cdots\right)\:,\quad\mu>M_{R}\,,\label{eq:KHbeta}
\end{equation}
where the ellipsis $\cdots$ represent SM gauge terms. Then, above
$M_{R}$, the neutrino Yukawa couplings drive the top quark Yukawa
coupling to diverge at some scale $\Lambda\sim10^{20}$ GeV if the
ratio $y_{\nu}/y_{t}$ is large enough.

What about the Higgs boson mass? In section \ref{subsec:The-SM-case}
we have seen that in the NJL scheme quartic couplings should also
diverge at the scale $\Lambda$. However, in the pure SM this leads
to a too heavy Higgs boson. Unfortunately, the introduction of the
neutrino Yukawa couplings does not help here, in fact it even worsens
the situation because Yukawa couplings give always a negative contribution
to the running of quartic couplings. KH solve this problem by introducing
a fundamental neutral scalar singlet, $S$, at the electroweak scale
like in scalar Higgs portal models \citep{Patt:2006fw,McDonald:1993ex,Silveira:1985rk}.
In these models, if the singlet scalar develops a VEV, $\left\langle S\right\rangle \gg\left\langle H\right\rangle $
the mass of the lighter scalar is given by 
\begin{equation}
m_{h}^{2}\simeq2\left(\lambda_{H}-\frac{\lambda_{HS}^{2}}{\lambda_{S}}\right)\left\langle H\right\rangle ^{2}\,,\label{eq:KHhiggsmass}
\end{equation}
where $\lambda_{H},\lambda_{S},\lambda_{HS}$ are the quartic couplings
of $H,$$S$ and mixed $H,$$S$ respectively. From Eq.~\eqref{eq:KHhiggsmass}
it is clear that $m_{h}$ can be relatively small even if the quartic
couplings are order one. Moreover, in the KH setup, only $\lambda_{H}$
is fixed by the compositeness boundary conditions, while $\lambda_{HS}$
and $\lambda_{S}$ are arbitrary and can be adjusted at will.

\section{The inverse see-saw model with composite scalars\label{sec:the-model}}

\subsection{The inverse see-saw model of neutrino masses\label{subsec:The-inverse-see-saw}}

A nice way to lower the see-saw scale at arbitrary scales is provided
by the so-called inverse see-saw (ISS) mechanism. In this mechanism
\citep{Mohapatra:1986aw,Mohapatra:1986bd} one introduces, in addition
to 3 right-handed neutrinos, $\nu_{R}$, 3 new singlet fermions $n_{L}$,
with Lagrangian
\begin{equation}
\mathcal{L}_{\mathrm{iss}}=\overline{L}_{L}y_{\nu}\nu_{R}H+\bar{\nu}_{R}M_{\nu n}n_{L}+\frac{1}{2}\overline{n_{L}^{c}}\mu_{n}n_{L}+\mathrm{h.c.}\,,\label{eq:issGenLagrangian}
\end{equation}
where $y_{\nu}$, $M_{\nu n}$ and $\mu_{n}$ are 3$\times3$ matrices.
Notice that, if $\mu_{n}=0$, lepton number can be assigned in such
a way that it is conserved. After SSB the Lagrangian Eq.~\eqref{eq:issGenLagrangian}
leads to the following Majorana mass term
\begin{equation}
\mathcal{L}_{\mathrm{iss}}=\frac{1}{2}\begin{pmatrix}\overline{\nu_{L}^{c}} & \overline{\nu_{R}} & \overline{n_{L}^{c}}\end{pmatrix}\begin{pmatrix}0 & y_{\nu}^{*}\left\langle H\right\rangle  & 0\\
y_{\nu}^{\dagger}\left\langle H\right\rangle  & 0 & M_{\nu n}\\
0 & M_{\nu n}^{T} & \mu_{n}
\end{pmatrix}\begin{pmatrix}\nu_{L}\\
\nu_{R}^{c}\\
n_{L}
\end{pmatrix}+\mathrm{h.c.}\label{eq:issMass}
\end{equation}
If $\mu_{n}=0$, this can be diagonalized exactly and leads to 3 Dirac
neutrinos, whose masses squared are the eigenvalues of the matrix
$M_{\nu_{H}}^{2}=y_{\nu}^{*}y_{\nu}\left\langle H\right\rangle ^{2}+M_{\nu n}^{\dagger}M_{\nu n}$,
and 3 exactly massless Weyl neutrinos. If $\mu_{n}\not=0$, lepton
number is explicitly broken and the would be massless neutrinos acquire
a mass matrix given by (in the limit $M_{\nu n}\gg y_{\nu}\left\langle H\right\rangle $)
\begin{equation}
m_{\nu}\simeq y_{\nu}^{*}\frac{\left\langle H\right\rangle }{M_{\nu n}^{T}}\mu_{n}\frac{\left\langle H\right\rangle }{M_{\nu n}}y_{\nu}^{\dagger}\,,\label{eq:iss-mnu}
\end{equation}
so that if $\mu_{n}$ is small $m_{\nu}$ can be below $1$ eV even
if $y_{\nu}$ is order one and $M_{\nu n}$ about $1$~TeV.

An interesting variation consists in taking $\mu_{n}=0$ and adding
a Majorana mass term for right-handed neutrinos, $\overline{\nu_{R}^{c}}\mu_{\nu}\nu_{R}$.
In that case, active neutrino masses are not generated at tree level
(the determinant of the mass matrix remains zero), but are generated
at one loop \citep{Dev:2012sg}. Neutrino masses are given by a similar
expression but with an extra loop suppression factor, which allows
for larger values of $\mu_{\nu}$.

\subsection{The inverse see-saw model with composite scalars\label{subsec:njl-iss-description}}

In the following we will embed this mechanism in the BHL scheme, and
the interesting thing is that, since the masses of the new neutral
heavy leptons could be naturally at the electroweak scale, they can
be obtained through SSB of a composite singlet scalar at low scales
and implement the Higgs portal mechanism to accommodate the Higgs
boson mass.

We will consider a Lagrangian with only fermions and the following
interactions and Majorana mass terms\footnote{For simplicity we use only one family of leptons and $n_{L}$, but
the mechanism can be generalized easily to 3 families à la KH. Moreover,
to generate masses for the other quarks and leptons one should also
introduce additional four-fermion interactions, which will be neglected
here.}
\begin{align}
\mathcal{L}_{4f} & =\frac{h_{\nu}^{2}}{m_{0H}^{2}}\left(\overline{L}_{L}\nu_{R}\right)\left(\overline{\nu}_{R}L_{L}\right)+\frac{h_{s}^{2}}{m_{0H}^{2}}\left(\overline{n}_{L}\nu_{R}\right)\left(\overline{\nu}_{R}n_{L}\right)\nonumber \\
 & +\left(\frac{h_{t\nu}^{2}}{m_{0H}^{2}}\left(\overline{L}_{L}\nu_{R}\right)\left(\overline{t}_{R}T_{L}\right)+\frac{1}{2}\overline{n_{L}^{c}}\mu_{n}n_{L}\mathrm{\,+\,h.c}.\right)\,,\label{eq:iss4fLagrangian}
\end{align}
where the Majorana mass term for $n_{L}$, ~$\mu_{n}$, can be included
because $n_{L}$ is a singlet and it is necessary to obtain masses
for active neutrinos (as discussed before an alternative would be
to add a right-handed neutrino Majorana mass term $\overline{\nu_{R}^{c}}\mu_{\nu}\nu_{R}$).
Notice that this Lagrangian, when $\mu_{n}=0$, preserves two global
phase symmetries, $L_{\nu_{R}}\,:\;\nu_{R}\rightarrow e^{i\alpha}\nu_{R}$,
$L_{L}\rightarrow e^{i\alpha}L_{L}$ and $L_{n_{L}}\,:\;n_{L}\rightarrow e^{i\beta}n_{L}$.
If $\mu_{n}\not=0$, $L_{n_{L}}$ is explicitly broken but $L_{\nu_{R}}$
is preserved; if $\mu_{\nu}\not=0$, $L_{\nu_{R}}$ would be broken
but $L_{n_{L}}$ would be preserved and, finally, a term $\overline{\nu}_{R}n_{L}$
would break the two but keep $L_{\nu_{L}}+L_{n_{L}}$. This Lagrangian
can be obtained (in the limit in which $h_{\nu}\gg h_{t\nu}$ ) from
\begin{equation}
\mathcal{L}_{\Lambda}=-m_{0H}^{2}H^{\dagger}H+y_{0t}\overline{T}_{L}t_{R}H+y_{0\nu}\overline{L}_{L}\nu_{R}H-m_{0S}^{2}S^{\dagger}S+y_{0s}S\overline{\nu}_{R}n_{L}+\frac{1}{2}\overline{n_{L}^{c}}\mu_{n}n_{L}+\mathrm{h.c.}\,,\label{eq:issLagrangianLambda}
\end{equation}
where $S$ is a singlet scalar field which will be interpreted as
a $\bar{n}_{L}\nu_{R}$ bound state. Fermion loops will induce a scalar
potential and kinetic terms for the scalars
\begin{align}
\mathcal{L}_{\mu} & =Z_{H}(\mu)|D_{\mu}H|^{2}-\tilde{m}_{H}^{2}(\mu)|H|^{2}+Z_{S}(\mu)|\partial_{\mu}S|^{2}-\tilde{m}_{S}^{2}(\mu)|S|^{2}\nonumber \\
 & -\frac{1}{2}\tilde{\lambda}_{H}(\mu)|H|^{4}-\frac{1}{2}\tilde{\lambda}_{S}(\mu)|S|^{4}-\frac{1}{2}\tilde{\lambda}_{HS}(\mu)|H|^{2}|S|^{2}\nonumber \\
 & +\left(\tilde{y}_{t}(\mu)\overline{T}_{L}t_{R}H+\tilde{y}_{\nu}(\mu)\overline{L}_{L}\nu_{R}H+\tilde{y}_{s}(\mu)S\overline{\nu}_{R}n_{L}+\frac{1}{2}\overline{n_{L}^{c}}\mu_{n}n_{L}+\mathrm{h.c.}\right)\,.\label{eq:issLagrangian-mupre}
\end{align}
Calculation of the corresponding fermion loops and imposing the ``compositeness''
boundary conditions
\[
Z_{H}(\Lambda)=Z_{S}(\Lambda)=0\,,\quad\tilde{\lambda}_{H}(\Lambda)=\tilde{\lambda}_{S}(\Lambda)=\tilde{\lambda}_{HS}(\Lambda)=0\,,
\]
\begin{equation}
\tilde{m}_{H}^{2}(\Lambda)=m_{0H}^{2}\:,\quad\tilde{m}_{S}^{2}(\Lambda)=m_{0S}^{2}\,,\;\tilde{y}_{t}(\Lambda)=y_{0t}\,,\;\tilde{y}_{\nu}(\Lambda)=y_{0\nu}\,,\;\tilde{y}_{s}(\Lambda)=y_{0s}\,,\label{eq:issboundaries1}
\end{equation}
gives
\[
Z_{H}(\mu)=\left(y_{0\nu}^{2}+N_{c}y_{0t}^{2}\right)L(\mu)\,,\quad Z_{S}(\mu)=y_{0s}^{2}L(\mu)\,,
\]
\[
\tilde{\lambda}_{H}(\mu)=\left(2y_{0\nu}^{4}+2N_{c}y_{0t}^{4}\right)L(\mu)\,,\quad\tilde{\lambda}_{S}(\mu)=2y_{0s}^{4}L(\mu)\,,\quad\tilde{\lambda}_{HS}(\mu)=2y_{0\nu}^{2}y_{0s}^{2}L(\mu)\,
\]
\begin{equation}
\tilde{m}_{H}^{2}(\mu)=m_{0H}^{2}-\left(2y_{0\nu}^{2}+2N_{c}y_{0t}^{2}\right)\frac{1}{16\pi^{2}}\left(\Lambda^{2}-\mu^{2}\right)\,,\quad\tilde{m}_{S}^{2}(\mu)=m_{0S}^{2}-\frac{y_{0s}^{2}}{8\pi^{2}}\left(\Lambda^{2}-\mu^{2}\right)\label{eq:boundaries2}
\end{equation}
and, as in the SM case, $\tilde{y}_{t}(\mu)=\tilde{y}_{t}(\Lambda)=y_{0t}$,
$\tilde{y}_{\nu}(\mu)=\tilde{y}_{\nu}(\Lambda)=y_{0\nu}$, $\tilde{y}_{s}(\mu)=\tilde{y}_{s}(\Lambda)=y_{0s}$.

Now one rescales the scalar fields $H\rightarrow H/\sqrt{Z_{H}(\mu)}$,
$S\rightarrow S/\sqrt{Z_{S}(\mu)}$ to obtain

\begin{align}
\mathcal{L}_{\mu R} & =|D_{\mu}H|^{2}-m_{H}^{2}(\mu)|H|^{2}+|\partial_{\mu}S|^{2}-m_{S}^{2}(\mu)|S|^{2}\nonumber \\
 & -\frac{1}{2}\lambda_{H}(\mu)|H|^{4}-\frac{1}{2}\lambda_{S}(\mu)|S|^{4}-\frac{1}{2}\lambda_{HS}(\mu)|H|^{2}|S|^{2}+\nonumber \\
 & +\left(y_{t}(\mu)\overline{T}_{L}t_{R}H+y_{\nu}(\mu)\overline{L}_{L}\nu_{R}H+y_{s}(\mu)S\overline{\nu}_{R}n_{L}+\frac{1}{2}\overline{n_{L}^{c}}\mu_{n}n_{L}+\mathrm{h.c.}\right)\,,\label{eq:issRLagrangian}
\end{align}

with 
\begin{equation}
m_{H}^{2}(\mu)=\tilde{m}_{H}^{2}(\mu)/Z_{H}(\mu)\:,\qquad m_{S}^{2}(\mu)=\tilde{m}_{S}^{2}(\mu)/Z_{S}(\mu)\,,\;y_{s}^{2}(\mu)=y_{0s}^{2}/Z_{S}(\mu)=\frac{1}{L(\mu)}\,,\label{eq:isspars1}
\end{equation}
\begin{equation}
y_{t}^{2}(\mu)=y_{0t}^{2}/Z_{H}(\mu)=\frac{p^{2}}{\left(1+N_{c}p^{2}\right)L(\mu)}\,,\;y_{\nu}^{2}(\mu)=y_{0\nu}^{2}/Z_{H}(\mu)=\frac{1}{\left(1+N_{c}p^{2}\right)L(\mu)}\,,\label{eq:isspars2}
\end{equation}
\begin{equation}
\lambda_{H}(\mu)=\tilde{\lambda}_{H}(\mu)/Z_{H}^{2}(\mu)=\frac{2\left(1+N_{c}p^{4}\right)}{\left(1+N_{c}p^{2}\right)^{2}L(\mu)}\:,\quad\lambda_{S}(\mu)=\tilde{\lambda}_{S}(\mu)/Z_{S}^{2}(\mu)=\frac{2}{L(\mu)}\:,\label{eq:isspars3}
\end{equation}
\begin{equation}
\lambda_{HS}(\mu)=\tilde{\lambda}_{HS}(\mu)/(Z_{H}(\mu)Z_{S}(\mu))=\frac{2}{\left(1+N_{c}p^{2}\right)L(\mu)}\:,\qquad p\equiv y_{0t}/y_{0\nu}\,,\label{eq:isspars4}
\end{equation}
where we have defined $p\equiv y_{0t}/y_{0\nu}$, which characterizes
the relative strength of top quark to neutrino interactions and must
be small.

If the two scalar fields develop a VEV, the model specified by the
Lagrangian in Eq.~\eqref{eq:issRLagrangian} implements the ISS mechanism
described in Section~\ref{sec:the-model} with a mass $M_{\nu n}=y_{s}\left\langle S\right\rangle $.
Therefore, if $\mu_{n}\ll\left\langle H\right\rangle \ll\left\langle S\right\rangle $
one can explain small neutrino masses. Moreover, with this hierarchy
of scales, one can also implement the Higgs portal model \citep{Patt:2006fw,McDonald:1993ex,Silveira:1985rk}
in which the effective low-energy Higgs quartic coupling, $\lambda$,
can be small even if the complete theory quartic couplings are large,
as usually required in NJL scenarios (see Section~\ref{subsec:The-Krog-Hill}).
This leads to the following hierarchy of masses: $m_{\nu}\sim\mu_{n}\left\langle H\right\rangle ^{2}/\left\langle S\right\rangle ^{2}\ll m_{t},m_{h}\propto\left\langle H\right\rangle \ll M\sim m_{s},m_{\nu_{H}}\propto\left\langle S\right\rangle \ll\Lambda$,
where we have denoted by $M$, generically, the scale of new particles,
the scalar $S$, $m_{s},$ and the neutral heavy leptons $\nu_{H}$,
$m_{\nu_{H}}$. Notice that since the scalar potential has an extra
global symmetry\footnote{This is just a consequence of the global symmetries of the new four-fermion
interactions we have introduced.}, $S\rightarrow e^{i\alpha}S$, broken spontaneously, the low energy
spectrum contains, in addition to the SM fields, a Goldstone boson
coupled mainly to the neutral heavy particles. Then, it is a kind
of singlet Majoron \citep{Chikashige:1980ui} (triplet and doublet
Majorons \citep{Gelmini:1980re,Bertolini:1987kz} are now excluded
because the well measured invisible decay width of the $Z$ boson).
The phenomenology of this type of models is very interesting and one
can usually cope with it (a detailed phenomenological study of the
model and some of its variations will be given elsewhere). Just mention
that the mixing of the singlet scalar with the doublet will induce
modifications of the Higgs boson couplings which are experimentally
constrained and an invisible decay of the Higgs boson to Majorons,
which is also constrained. These constraints can be satisfied by taking
$\left\langle S\right\rangle $ large enough. Here we are more interested
in the possibility of obtaining the observed top quark and the Higgs
boson masses in this NJL scenario. Since the Majorana mass for the
new fermion $n_{L}$ must be much below the electroweak scale, $\mu_{n}\ll\left\langle H\right\rangle $,
it will not affect this calculation and can be safely neglected. We
will reintroduce it at the end when we discuss neutrino masses.

\subsection{The top quark and Higgs boson masses\label{subsec:top-higgs-masses}}

To obtain the value of $m_{t}$ we take the measured values of the
gauge couplings at $m_{Z}$, and run them up to $\Lambda_{\kappa}$
with SM RGE. Since the new particles are all singlets, at one loop,
they do not affect the running of gauge couplings. At the scale $\Lambda_{\kappa}$
we impose the boundary conditions Eqs.~(\ref{eq:isspars1}--\ref{eq:isspars4})
and obtain all Yukawa couplings, $y_{t}$, $y_{\nu}$, $y_{s}$, and
quartic couplings $\lambda_{H},\lambda_{S},\lambda_{HS}$, as functions
of $\kappa$ and $p$. Then we run them with the RGE of the complete
model (see the Appendix for the beta functions) up to the scale $M$,
which we fix at some value above the electroweak scale. At the scale
$M$ we assume that $S$ develops a VEV\footnote{We assume that the parameters of the model are adjusted such as both,
the doublet $H$ and the singlet $S$, develop a vacuum expectation
value, i.e. $m_{H}^{2}(M)<0$ and $m_{S}^{2}(M)<0$.} giving a $\bar{\nu}_{R}n_{L}$ mass term so that the fermions $\nu_{R}$
and $n_{L}$ combine to form a Dirac fermion (if $\mu_{n}=0$, if
$\mu_{n}\not=0$ a pseudo-Dirac fermion) with mass $m_{\nu_{H}}\sim y_{s}(M)\left\langle S\right\rangle \sim M$.
Then, to obtain the top quark mass, we decouple the heavy particles
and run the top quark Yukawa coupling with the SM RGE from $M$ to
$\mu=m_{t}$, which at this point is still unknown but can easily
be computed by using the SM relation\footnote{Therefore, we stop the running when this equation is satisfied.}
\begin{equation}
y_{t}(m_{t})=\sqrt{2}\frac{m_{t}}{v}\left(1+\delta_{t}\right)\quad,\qquad\delta_{t}\approx-0.059\,.\label{eq:top-to-yukawa-relation}
\end{equation}
Here $\delta_{t}$ represents the well known SM corrections to the
relation between the top quark pole mass and the Yukawa coupling~\citep{Kniehl:2014yia,Hempfling:1994ar}.
$\delta_{t}$ includes QCD corrections which, as commented in Section~\ref{sec:The-BHL-mechanism},
are very large, and some small electroweak corrections. For masses
$m_{t}\sim173$~GeV and $m_{h}\sim125$~GeV, $\delta_{t}$ can be
well approximated by the number given above, which we use in the following
calculations, but it can be computed for arbitrary values of $m_{t}$
and $m_{h}$.

\begin{figure}[h]
\begin{centering}
\includegraphics[scale=0.48]{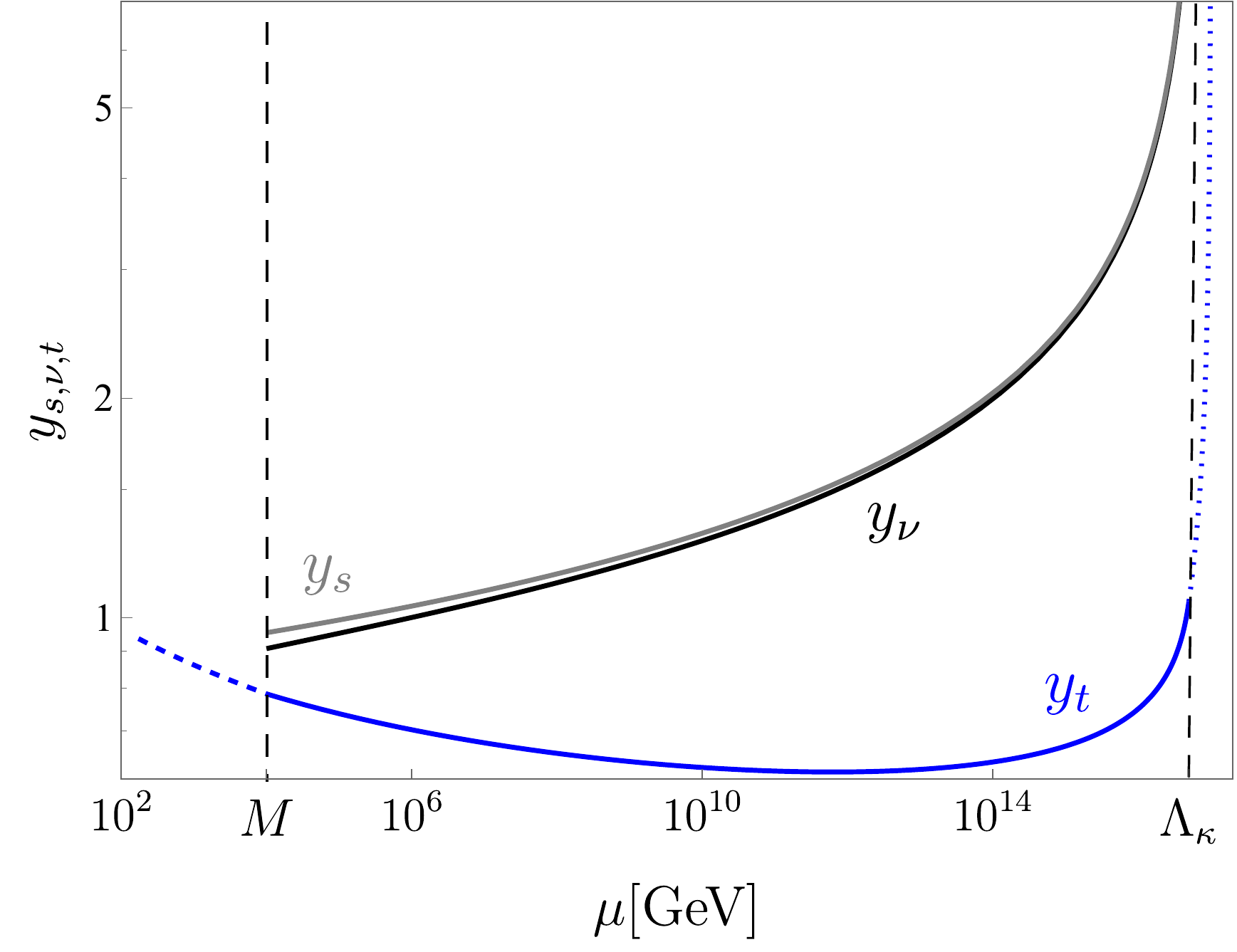}
\par\end{centering}
\caption{Evolution of the Yukawa couplings, as explained in the text, for $p=0.1$,
$\Lambda=10^{17}$~GeV, $\kappa=2$. The new heavy particles are
assumed to have a mass $M=10$~TeV where they decouple. Then from
$M$ to the electroweak scale the top quark Yukawa coupling $y_{t}$
is run according to SM RGE. For the chosen values this procedure gives
finally $m_{t}=173$~GeV.\label{fig:iss-yukawa-run}}
\end{figure}

We represent in Fig.~\ref{fig:iss-yukawa-run} an example of the
running of all Yukawa couplings for $p=0.1$, $\Lambda=10^{17}$~GeV,
$\kappa=2$ and $M=10$~TeV that reproduces the correct value of
$m_{t}\sim173$~GeV. The SM RGE running is shown in the dashed-blue
line, while the running with the new particles is shown in the solid
blue line (from $M$ to $\Lambda_{\kappa}$). Above $\Lambda_{\kappa}$
all Yukawa couplings run only with fermion loops -dotted line- and
meet the Landau pole at $\mu=\Lambda.$ We see how the Yukawa couplings
$y_{\nu}$ and $y_{s}$ pull the top quark Yukawa coupling, $y_{t}$,
towards the Landau pole. 

To obtain the Higgs boson mass we have to study the Higgs potential.
As commented before we assume that the two scalars obtain a VEV, then
we write
\begin{equation}
H^{(0)}=\frac{1}{\sqrt{2}}\left(v+h+i\omega^{(0)}\right)\:,\quad S=\frac{1}{\sqrt{2}}\left(u+s+i\theta\right)\,.\label{eq:ScalarExpansion}
\end{equation}
Since the potential has an extra global symmetry, $S\rightarrow e^{i\alpha}S$,
broken spontaneously, the low energy spectrum (below $\mu=M$) contains,
in addition to the SM fields, a Goldstone boson, which is given by
the imaginary part of $S$, $\theta$. On the other hand the real
part of $S$ mixes with the Higgs doublet with a mass matrix squared
given by (in the $(h,s$) basis)
\begin{equation}
M_{\mathrm{scalars}}^{2}=\begin{pmatrix}\lambda_{H}v^{2} & \lambda_{HS}vu\\
\lambda_{HS}vu & \lambda_{S}u^{2}
\end{pmatrix}\:,\quad v\equiv\sqrt{2}\left\langle H\right\rangle \:,\quad u\equiv\sqrt{2}\left\langle S\right\rangle \,.\label{eq:ScalarMassMatrix}
\end{equation}
The smallest of the eigenvalues, $m_{h}^{2}$, can be identified with
the observed Higgs boson mass squared while the largest will give
the mass squared of the new scalar (for $u\gg v$, $m_{s}^{2}\sim\lambda_{S}u^{2}$).
It is easy to check that these two eigenvalues are related by 
\[
m_{h}^{2}=v^{2}\left(\lambda_{H}-\frac{\lambda_{HS}^{2}}{\lambda_{S}}\right)\frac{1-\lambda_{H}v^{2}/m_{s}^{2}}{1-\left(\lambda_{H}-\lambda_{HS}^{2}/\lambda_{S}\right)v^{2}/m_{s}^{2}}\stackrel{m_{s}\gg v}{\longrightarrow}
\]
\begin{equation}
\stackrel{m_{s}\gg v}{\longrightarrow}v^{2}\left(\lambda_{H}-\frac{\lambda_{HS}^{2}}{\lambda_{S}}\right)\left(1-\frac{\lambda_{HS}^{2}}{\lambda_{S}}\frac{v^{2}}{m_{s}^{2}}+\cdots\right)\,.\label{eq:mhcorrectionatM}
\end{equation}
Then if $m_{s}\gg v$, the effect of the new scalar on the Higgs boson
mass is just a redefinition of the SM quartic coupling, $\lambda$,
in terms of the couplings of the complete theory\footnote{We perform the matching at a scale $M$ of the order of the mass of
the new particles, the fermions and the scalars, which we assume are
of the same order. Since for $u\gg v,$ $m_{\nu_{H}}\text{\ensuremath{\sim y_{s}u/\sqrt{2}}}$
and $m_{s}^{2}\sim\lambda_{S}u^{2}$, one needs $\lambda_{S}$ and
$y_{s}^{2}$ to be of the same order. This is guaranteed by the boundary
conditions at the $\Lambda_{\kappa}$ scale. To be definite, in our
calculations we take $M=m_{s}$, but we have checked that this condition
is not strongly modified by the running from $\Lambda_{\kappa}$ to
$M$.} 
\begin{equation}
\lambda(M)=\left(\lambda_{H}(M)-\frac{\lambda_{HS}^{2}(M)}{\lambda_{S}(M)}\right)\:,\label{eq:lambda-matching-M}
\end{equation}
with corrections, $\delta_{hs},$ which vanish for $m_{s}\gg v$.
Some electroweak corrections can be incorporated by running $\lambda$
from $M$ to $m_{t}$ according to the SM. Finally, to connect $\lambda(m_{t})$
with the physical Higgs boson mass, $m_{h}$, one should also take
into account the well known SM corrections~\citep{Sirlin:1985ux},
$\delta_{h}$. Thus, one has

\begin{equation}
m_{h}^{2}=\lambda(m_{t})v^{2}\frac{1+\delta_{hs}}{1+\delta_{h}}\:,\quad\delta_{h}\sim-0.011\,,\label{eq:mh2mass}
\end{equation}
where $\delta_{h}$ is given by a complicated expression that depends
on the masses of SM particles \citep{Sirlin:1985ux}, but for $m_{h}\sim125$~GeV
and $m_{t}\sim173$~GeV it is well approximated by the value above,
while $\delta_{hs}$ is obtained from Eq.~\eqref{eq:mhcorrectionatM}

\begin{equation}
\delta_{hs}=-\frac{\lambda_{HS}^{2}(M)}{\lambda_{S}(M)}\frac{v^{2}}{m_{s}^{2}}\left(1-\left(\lambda_{H}(M)-\frac{\lambda_{HS}^{2}(M)}{\lambda_{S}(M)}\right)\frac{v^{2}}{m_{s}^{2}}\right)^{-1}\,.\label{eq:deltahs}
\end{equation}

To evaluate these expressions we need the $\lambda_{H,S,SH}$ couplings
at the scale $M$ and $\lambda$ at the scale $m_{t}$. For that we
run them using the beta functions given in the Appendix. In Fig.~\ref{fig:iss-lamdas-run}
we give an example with the same values of $p$, $M,$$\Lambda$ and
$\kappa$ as in Fig.~\ref{fig:iss-yukawa-run}, where we see how
$\lambda_{H,S,SH}$ evolve to lower energies from the Landau pole.
Since at $\mu=M=10$~TeV all the heavy particles decouple, the couplings
$\lambda_{H,S,SH}$ do not run anymore but leave a SM-like theory
with an effective coupling $\lambda(M)$ given by Eq.~\eqref{eq:lambda-matching-M}.
Then, from $M$ to the electroweak scale $\lambda$ runs according
the SM RGE. This procedure gives finally (for the chosen values of
$p$, $M,$$\Lambda$ and $\kappa$) $m_{h}=125$~GeV (and $m_{t}=173\,$GeV).

\begin{figure}[h]
\begin{centering}
\includegraphics[scale=0.48]{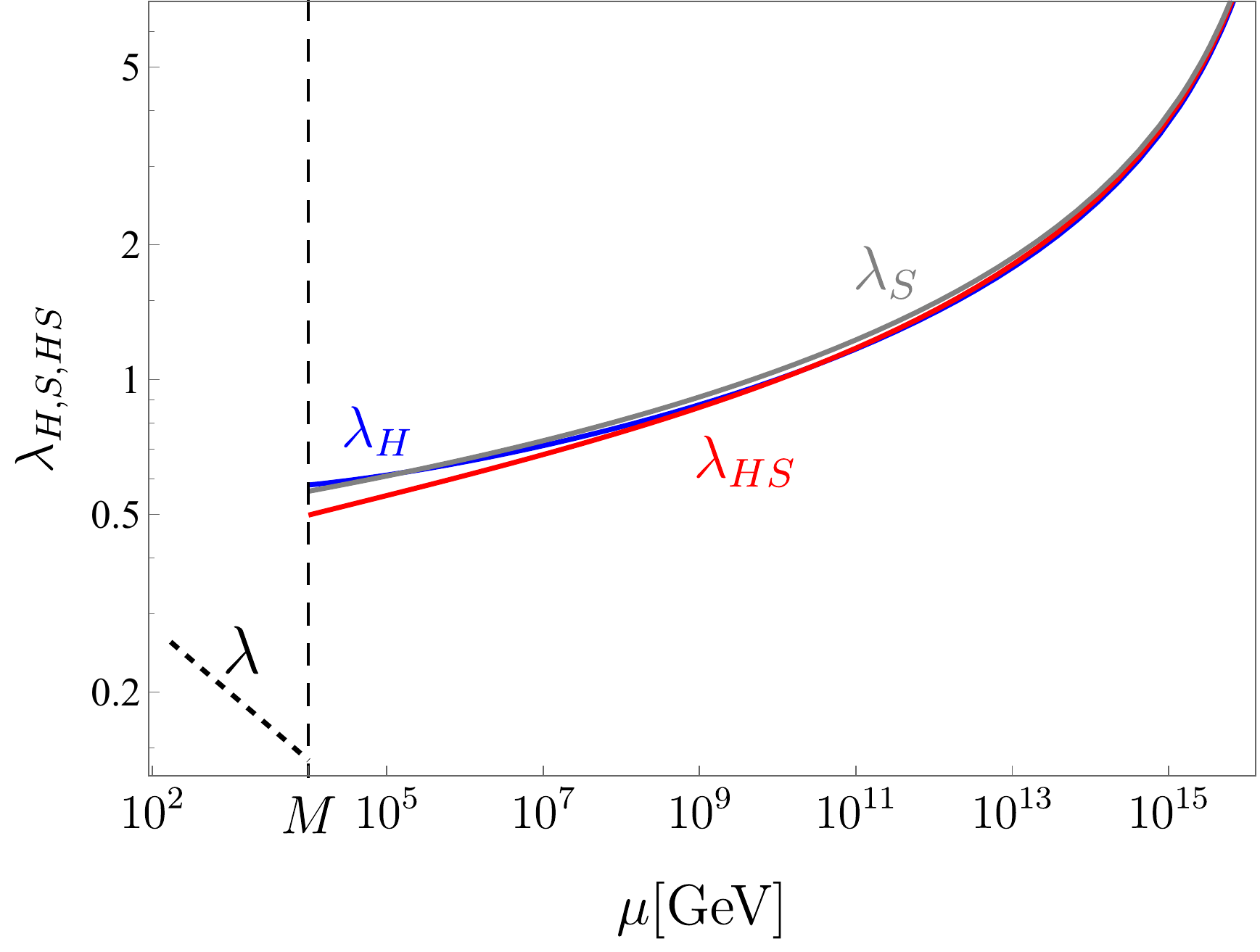}
\par\end{centering}
\caption{Evolution of the scalar quartic couplings, as explained in the text,
for the same values as in Fig.~\ref{fig:iss-yukawa-run}. At the
scale $M=10$ TeV the new particles decouple leaving a SM quartic
coupling, given by Eq.~\eqref{eq:lambda-matching-M}, which runs
up to the weak scale according to the SM RGE. For the chosen values
this procedure gives finally $m_{h}=125\,$GeV.\label{fig:iss-lamdas-run}}
\end{figure}

We can repeat this procedure for different values of $p,\Lambda$
(and $\kappa$, $M$) and check if they are able to reproduce the
measured values of $m_{t}\sim173\,$GeV and $m_{h}\sim125$~GeV.

\begin{figure}[h]
\begin{centering}
\includegraphics[scale=0.48]{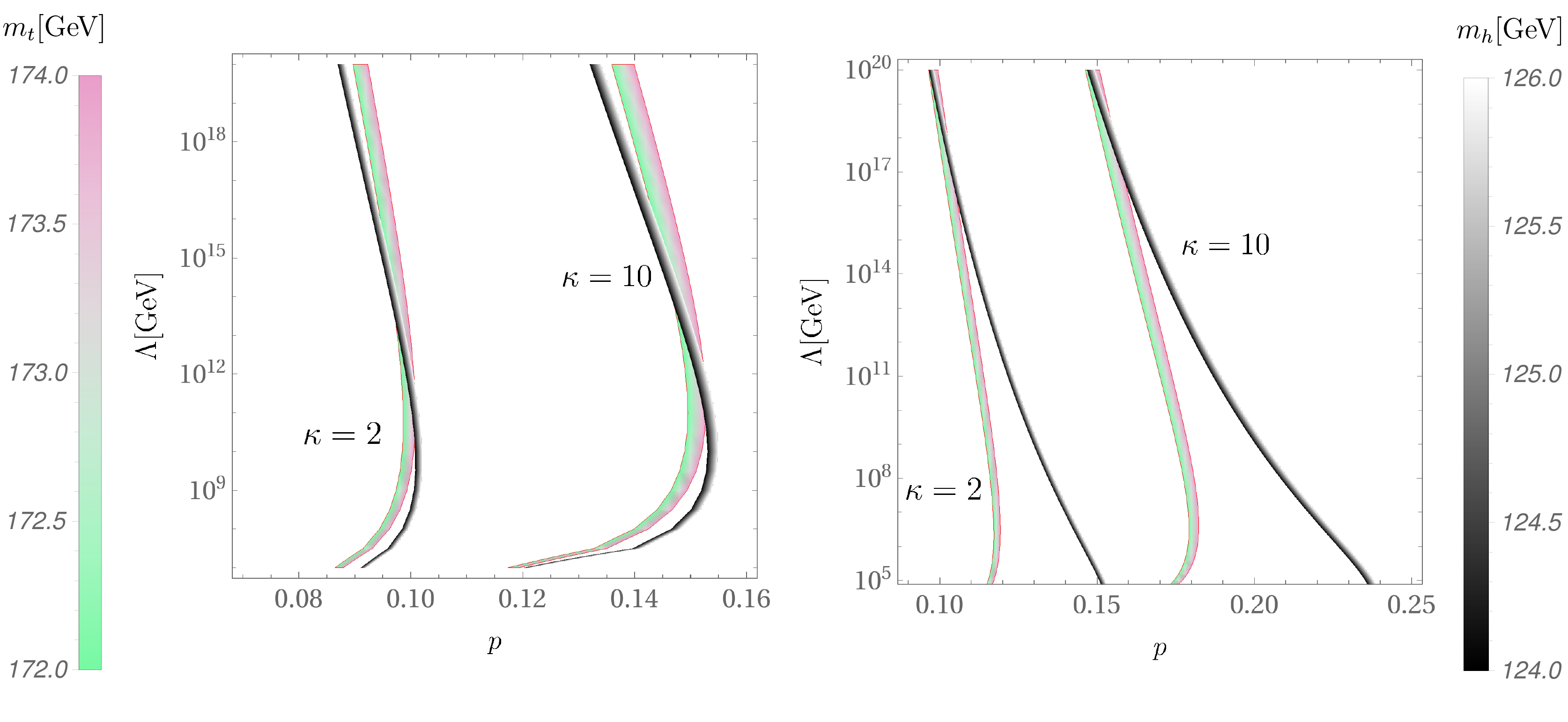}
\par\end{centering}
\caption{Region of $p,\Lambda$ that can reproduce values of $m_{t}$ in a
region of $1$~GeV around $m_{t}=173$ GeV (band with green-pink
colors) and $m_{h}$ in a region of 1~GeV around $m_{h}=125$~GeV
(gray band). On the left for $M=1000\,$TeV and on the right for  $M=1\,$TeV.
In each plot we present results for two different values of $\kappa$.\label{fig:densityplots}}
\end{figure}

In Fig.~\ref{fig:densityplots} we depict the region of $p,\Lambda$
that can reproduce values of $m_{t}$ in a region of $1$~GeV around
$m_{t}=173$ GeV (band with green-pink colors) and $m_{h}$ in a region
of 1~GeV around $m_{h}=125$~GeV (gray band). We do this for two
values of $\kappa$ in each plot (for fixed $M=1\,$TeV on the right
and $M=1000\,$TeV on the left). We see that, indeed, there is an
overlapping region where one can reproduce both the Higgs boson and
the top quark masses. For $M=1$ TeV this is found around $\Lambda\sim10^{19}$~GeV,
while for $M=1000\,$TeV the overlapping region occurs around $\Lambda\sim10^{12}$
GeV. Larger values of $M$ lead to lower values of $\Lambda$ but
for $M\gtrsim10^{8}$ GeV there are no solutions. The effect of the
exact scale at which we perform the matching between the complete
model and the model with static scalars, which is parametrized by
$\kappa=\Lambda/\Lambda_{\kappa}$, only changed the preferred value
of $p$, which is always small, as required for consistency. We only
represent values of $\Lambda$ a couple of orders of magnitude above
$M$, since for $\Lambda$ close to $M$ the range of running is very
small and the results are completely dominated by the matching, which
cannot reliably be computed without knowing the details of the complete
theory behind the four-fermion interactions.

Once $m_{t}$ and $m_{h}$ are obtained with the correct values, all
the couplings and scales are quite constrained. However, the Majorana
mass terms of the heavy fermions, $\mu_{\nu}$ and/or $\mu_{n}$,
are completely free and can be adjusted to obtain neutrino masses
below $1\,$eV using the inverse see-saw formula, Eq.~\eqref{eq:iss-mnu}.
A complete analysis of neutrino masses, as for the rest of fermions,
requires a three family analysis, but it is clear that given the freedom
in $\mu_{\nu,n}$ there should be no problem for adjusting neutrino
masses and mixings. Alternatively, one could also try to generate
the Majorana mass terms by using composite scalars breaking lepton
number as done in \citep{Dib:2019iqo} with the interesting consequences
discussed there.

\section{Conclusions\label{sec:Conclusions}}

Following previous work, Ref.~\citep{Krog:2015cna,Smetana:2013hm,Antusch:2002xh,Martin:1991xw},
we have explored the possibility that the observed Higgs boson is
mainly a bound state of neutrinos formed because a strong four-fermion
interaction between neutrinos appears at high scales. The minimal
version of this scenario has problems to reproduce the observed top
quark mass and, especially, the Higgs boson mass. We have overcome
these problems by introducing, in addition to right-handed neutrinos,
$\nu_{R}$, a new singlet fermion, $n_{L}$, with four-fermion interactions
$\left(\bar{\nu}_{R}n_{L}\right)\left(\bar{n}_{L}\nu_{R}\right)$,
which gives rise to a new scalar bound state. This singlet scalar
develops a VEV and, therefore, mixes with the Higgs doublet allowing
us to obtain a small Higgs mass even if the couplings are large, as
required in composite scalar models.

The compositeness condition basically fixes all Yukawa and quartic
couplings at the compositeness scale, therefore, the parameters of
the model are very constrained. In spite of that, this setup can accommodate
the correct masses for the top quark and Higgs boson for compositeness
scales below the Planck scale and masses of the new particles above
the electroweak scale but below $\sim10^{8}$~GeV. 

If small Majorana masses are allowed for $\nu_{R}$ and/or $n_{L}$,
we naturally obtain a low-scale see-saw scenario for neutrino masses
with the presence of additional neutral scalars coupled to the neutral
fermions. If the scale of the new particles is not much larger than
$1$ TeV the model exhibits a very rich phenomenology that could be
tested in present and near future experiments and will be studied
in another publication. Further extensions in which the Majorana mass
terms for $\nu_{R}$ and or $n_{L}$ are also generated by dynamical
symmetry breaking might also be interesting.

\begin{acknowledgments}
This work is partially supported by the FEDER/MCIyU-AEI grant FPA2017-84543-P,
by the ``Severo Ochoa'' Excellence Program under grant SEV-2014-0398
and by the \textquotedblleft Generalitat Valenciana\textquotedblright{}
under grant PROMETEO/2019/087. L.C and C.F. are also supported by
the ``Generalitat Valenciana'' under the ``GRISOLIA'' and ``ACIF''
fellowship programs, respectively.
\end{acknowledgments}

\appendix

\section{RGE of the model\label{sec:appendix}}

Here we give the RGE beta functions of the model, which have been
computed with the help of SARAH \citep{Staub:2013tta} (we use the
SU(5) convention $3g_{1}^{2}=5g^{\prime2}$ for the U(1) factor)

Gauge couplings (same as in the SM):
\begin{equation}
\beta_{g_{1}}=\frac{41}{10}g_{1}^{3}\;,\qquad\beta_{g_{2}}=-\frac{19}{6}g_{2}^{3}\;,\qquad\beta_{g_{3}}=-7g_{3}^{3}\label{eq:RGEgauge}
\end{equation}

Yukawas:
\begin{align}
\beta_{y_{t}} & =y_{t}\left(\frac{9}{2}y_{t}^{2}-8g_{3}^{2}-\frac{9}{4}g_{2}^{2}-\frac{17}{20}g_{1}^{2}+y_{\nu}^{2}\right)\nonumber \\
\beta_{y_{\nu}} & =y_{\nu}\left(\frac{5}{2}y_{\nu}^{2}+\frac{1}{2}y_{s}^{2}+3y_{t}^{2}-\frac{9}{20}\left(5g_{2}^{2}+g_{1}^{2}\right)\right)\label{eq:RGEYukawas}\\
\beta_{y_{s}} & =y_{s}\left(2y_{s}^{2}+y_{\nu}^{2}\right)\nonumber 
\end{align}

Quartic couplings:
\begin{align}
\beta_{\lambda_{H}} & =12\lambda_{H}^{2}+\frac{27}{100}g_{1}^{4}+\frac{9}{10}g_{1}^{2}g_{2}^{2}+\frac{9}{4}g_{2}^{4}+\lambda_{H}\left(12y_{t}^{2}-\frac{9}{5}g_{1}^{2}-9g_{2}^{2}+4y_{\nu}^{2}\right)+2\lambda_{HS}^{2}-4y_{\nu}^{4}-12y_{t}^{4}\nonumber \\
\beta_{\lambda_{S}} & =10\lambda_{S}^{2}+4\lambda_{S}y_{s}^{2}+4\lambda_{HS}^{2}-4y_{s}^{4}\label{eq:RGElambdas}\\
\beta_{\lambda_{HS}} & =\lambda_{HS}\left(4\lambda_{HS}+6\lambda_{H}+4\lambda_{S}+2y_{\nu}^{2}+2y_{s}^{2}+6y_{t}^{2}-\frac{9}{10}g_{1}^{2}-\frac{9}{2}g_{2}^{2}\right)-4y_{s}^{2}y_{\nu}^{2}\nonumber 
\end{align}


\begin{thebibliography}{10}

\bibitem{Bardeen:1989ds}
W.~A. Bardeen, C.~T. Hill and M.~Lindner, \emph{{Minimal Dynamical Symmetry
  Breaking of the Standard Model}},
  \href{https://doi.org/10.1103/PhysRevD.41.1647}{\emph{Phys. Rev.} {\bfseries
  D41} (1990) 1647}.

\bibitem{Nambu:1961fr}
Y.~Nambu and G.~Jona-Lasinio, \emph{{DYNAMICAL MODEL OF ELEMENTARY PARTICLES
  BASED ON AN ANALOGY WITH SUPERCONDUCTIVITY. II}},
  \href{https://doi.org/10.1103/PhysRev.124.246}{\emph{Phys. Rev.} {\bfseries
  124} (1961) 246}.

\bibitem{Nambu:1961tp}
Y.~Nambu and G.~Jona-Lasinio, \emph{{Dynamical Model of Elementary Particles
  Based on an Analogy with Superconductivity. 1.}},
  \href{https://doi.org/10.1103/PhysRev.122.345}{\emph{Phys. Rev.} {\bfseries
  122} (1961) 345}.

\bibitem{Miransky:1988xi}
V.~A. Miransky, M.~Tanabashi and K.~Yamawaki, \emph{{Dynamical Electroweak
  Symmetry Breaking with Large Anomalous Dimension and t Quark Condensate}},
  \href{https://doi.org/10.1016/0370-2693(89)91494-9}{\emph{Phys. Lett.}
  {\bfseries B221} (1989) 177}.

\bibitem{Miransky:1989ds}
V.~A. Miransky, M.~Tanabashi and K.~Yamawaki, \emph{{Is the t Quark Responsible
  for the Mass of W and Z Bosons?}},
  \href{https://doi.org/10.1142/S0217732389001210}{\emph{Mod. Phys. Lett.}
  {\bfseries A4} (1989) 1043}.

\bibitem{Suzuki:1989nv}
M.~Suzuki, \emph{{Composite Higgs Bosons in the {Nambu-Jona-Lasinio} Model}},
  \href{https://doi.org/10.1103/PhysRevD.41.3457}{\emph{Phys. Rev.} {\bfseries
  D41} (1990) 3457}.

\bibitem{Suzuki:1989si}
M.~Suzuki, \emph{{Formation of Composite Higgs Bosons From Quark - Anti-quarks
  at Lower Energy Scales}},
  \href{https://doi.org/10.1142/S0217732390001359}{\emph{Mod. Phys. Lett.}
  {\bfseries A5} (1990) 1205}.

\bibitem{Marciano:1989mj}
W.~J. Marciano, \emph{{Dynamical Symmetry Breaking and the Top Quark Mass}},
  \href{https://doi.org/10.1103/PhysRevD.41.219}{\emph{Phys. Rev.} {\bfseries
  D41} (1990) 219}.

\bibitem{Marciano:1989xd}
W.~J. Marciano, \emph{{HEAVY TOP QUARK MASS PREDICTIONS}},
  \href{https://doi.org/10.1103/PhysRevLett.62.2793}{\emph{Phys. Rev. Lett.}
  {\bfseries 62} (1989) 2793}.

\bibitem{Hill:1980sq}
C.~T. Hill, \emph{{Quark and Lepton Masses from Renormalization Group Fixe d
  Points}}, \href{https://doi.org/10.1103/PhysRevD.24.691}{\emph{Phys. Rev.}
  {\bfseries D24} (1981) 691}.

\bibitem{Hill:1985tg}
C.~T. Hill, C.~N. Leung and S.~Rao, \emph{{Renormalization Group Fixed Points
  and the Higgs Boson Spectrum}},
  \href{https://doi.org/10.1016/0550-3213(85)90328-1}{\emph{Nucl. Phys.}
  {\bfseries B262} (1985) 517}.

\bibitem{Cvetic:1997eb}
G.~Cvetic, \emph{{Top quark condensation}},
  \href{https://doi.org/10.1103/RevModPhys.71.513}{\emph{Rev. Mod. Phys.}
  {\bfseries 71} (1999) 513}
  [\href{https://arxiv.org/abs/hep-ph/9702381}{{\ttfamily hep-ph/9702381}}].

\bibitem{Hill:2002ap}
C.~T. Hill and E.~H. Simmons, \emph{{Strong dynamics and electroweak symmetry
  breaking}}, \href{https://doi.org/10.1016/S0370-1573(03)00140-6}{\emph{Phys.
  Rept.} {\bfseries 381} (2003) 235}
  [\href{https://arxiv.org/abs/hep-ph/0203079}{{\ttfamily hep-ph/0203079}}].

\bibitem{Krog:2015cna}
J.~Krog and C.~T. Hill, \emph{{Is the Higgs Boson Composed of Neutrinos?}},
  \href{https://doi.org/10.1103/PhysRevD.92.093005}{\emph{Phys. Rev.}
  {\bfseries D92} (2015) 093005}
  [\href{https://arxiv.org/abs/1506.02843}{{\ttfamily 1506.02843}}].

\bibitem{Smetana:2013hm}
A.~Smetana, \emph{{Top-quark and neutrino composite Higgs bosons}},
  \href{https://doi.org/10.1140/epjc/s10052-013-2513-8}{\emph{Eur. Phys. J.}
  {\bfseries C73} (2013) 2513}
  [\href{https://arxiv.org/abs/1301.1554}{{\ttfamily 1301.1554}}].

\bibitem{Antusch:2002xh}
S.~Antusch, J.~Kersten, M.~Lindner and M.~Ratz, \emph{{Dynamical electroweak
  symmetry breaking by a neutrino condensate}},
  \href{https://doi.org/10.1016/S0550-3213(03)00188-3}{\emph{Nucl. Phys.}
  {\bfseries B658} (2003) 203}
  [\href{https://arxiv.org/abs/hep-ph/0211385}{{\ttfamily hep-ph/0211385}}].

\bibitem{Martin:1991xw}
S.~P. Martin, \emph{{Dynamical electroweak symmetry breaking with top quark and
  neutrino condensates}},
  \href{https://doi.org/10.1103/PhysRevD.44.2892}{\emph{Phys. Rev.} {\bfseries
  D44} (1991) 2892}.

\bibitem{Dib:2019jod}
C.~Dib, S.~Kovalenko, I.~Schmidt and A.~Smetana, \emph{{Low-scale seesaw from
  neutrino condensation}},  \href{https://arxiv.org/abs/1904.06280}{{\ttfamily
  1904.06280}}.

\bibitem{Barenboim:2016xhn}
G.~Barenboim and C.~Bosch, \emph{{Composite states of two right-handed
  neutrinos}}, \href{https://doi.org/10.1103/PhysRevD.94.116019}{\emph{Phys.
  Rev.} {\bfseries D94} (2016) 116019}
  [\href{https://arxiv.org/abs/1610.06588}{{\ttfamily 1610.06588}}].

\bibitem{Barenboim:2010nm}
G.~Barenboim and J.~Rasero, \emph{{Baryogenesis from a right-handed neutrino
  condensate}}, \href{https://doi.org/10.1007/JHEP03(2011)097}{\emph{JHEP}
  {\bfseries 03} (2011) 097} [\href{https://arxiv.org/abs/1009.3024}{{\ttfamily
  1009.3024}}].

\bibitem{Barenboim:2008ds}
G.~Barenboim, \emph{{Inflation might be caused by the right: Handed neutrino}},
  \href{https://doi.org/10.1088/1126-6708/2009/03/102}{\emph{JHEP} {\bfseries
  03} (2009) 102} [\href{https://arxiv.org/abs/0811.2998}{{\ttfamily
  0811.2998}}].

\bibitem{Mohapatra:1986aw}
R.~N. Mohapatra, \emph{{Mechanism for Understanding Small Neutrino Mass in
  Superstring Theories}},
  \href{https://doi.org/10.1103/PhysRevLett.56.561}{\emph{Phys. Rev. Lett.}
  {\bfseries 56} (1986) 561}.

\bibitem{Mohapatra:1986bd}
R.~N. Mohapatra and J.~W.~F. Valle, \emph{{Neutrino Mass and Baryon Number
  Nonconservation in Superstring Models}},
  \href{https://doi.org/10.1103/PhysRevD.34.1642}{\emph{Phys. Rev.} {\bfseries
  D34} (1986) 1642}.

\bibitem{Wyler:1982dd}
D.~Wyler and L.~Wolfenstein, \emph{{Massless Neutrinos in Left-Right Symmetric
  Models}}, \href{https://doi.org/10.1016/0550-3213(83)90482-0}{\emph{Nucl.
  Phys.} {\bfseries B218} (1983) 205}.

\bibitem{Bernabeu:1987gr}
J.~Bernabeu, A.~Santamaria, J.~Vidal, A.~Mendez and J.~W.~F. Valle,
  \emph{{Lepton Flavor Nonconservation at High-Energies in a Superstring
  Inspired Standard Model}},
  \href{https://doi.org/10.1016/0370-2693(87)91100-2}{\emph{Phys. Lett.}
  {\bfseries B187} (1987) 303}.

\bibitem{Kniehl:2014yia}
B.~A. Kniehl and O.~L. Veretin, \emph{{Two-loop electroweak threshold
  corrections to the botto m and top Yukawa couplings}},
  \href{https://doi.org/10.1016/j.nuclphysb.2015.02.012,
  10.1016/j.nuclphysb.2014.05.029}{\emph{Nucl. Phys.} {\bfseries B885} (2014)
  459} [\href{https://arxiv.org/abs/1401.1844}{{\ttfamily 1401.1844}}].

\bibitem{Hempfling:1994ar}
R.~Hempfling and B.~A. Kniehl, \emph{{On the relation between the fermion pole
  mass and MS Yukawa coupling in the standard model}},
  \href{https://doi.org/10.1103/PhysRevD.51.1386}{\emph{Phys. Rev.} {\bfseries
  D51} (1995) 1386} [\href{https://arxiv.org/abs/hep-ph/9408313}{{\ttfamily
  hep-ph/9408313}}].

\bibitem{Patt:2006fw}
B.~Patt and F.~Wilczek, \emph{{Higgs-field portal into hidden sectors}},
  \href{https://arxiv.org/abs/hep-ph/0605188}{{\ttfamily hep-ph/0605188}}.

\bibitem{McDonald:1993ex}
J.~McDonald, \emph{{Gauge singlet scalars as cold dark matter}},
  \href{https://doi.org/10.1103/PhysRevD.50.3637}{\emph{Phys. Rev.} {\bfseries
  D50} (1994) 3637} [\href{https://arxiv.org/abs/hep-ph/0702143}{{\ttfamily
  hep-ph/0702143}}].

\bibitem{Silveira:1985rk}
V.~Silveira and A.~Zee, \emph{{SCALAR PHANTOMS}},
  \href{https://doi.org/10.1016/0370-2693(85)90624-0}{\emph{Phys. Lett.}
  {\bfseries 161B} (1985) 136}.

\bibitem{Dev:2012sg}
P.~S.~B. Dev and A.~Pilaftsis, \emph{{Minimal Radiative Neutrino Mass Mechanism
  for Inverse Seesaw Models}},
  \href{https://doi.org/10.1103/PhysRevD.86.113001}{\emph{Phys. Rev.}
  {\bfseries D86} (2012) 113001}
  [\href{https://arxiv.org/abs/1209.4051}{{\ttfamily 1209.4051}}].

\bibitem{Chikashige:1980ui}
Y.~Chikashige, R.~N. Mohapatra and R.~D. Peccei, \emph{{Are There Real
  Goldstone Bosons Associated with Broken Lepton Number?}},
  \href{https://doi.org/10.1016/0370-2693(81)90011-3}{\emph{Phys. Lett.}
  {\bfseries 98B} (1981) 265}.

\bibitem{Gelmini:1980re}
G.~B. Gelmini and M.~Roncadelli, \emph{{Left-Handed Neutrino Mass Scale and
  Spontaneously Broke n Lepton Number}},
  \href{https://doi.org/10.1016/0370-2693(81)90559-1}{\emph{Phys. Lett.}
  {\bfseries 99B} (1981) 411}.

\bibitem{Bertolini:1987kz}
S.~Bertolini and A.~Santamaria, \emph{{The Doublet Majoron Model and Solar
  Neutrino Oscillations}},
  \href{https://doi.org/10.1016/0550-3213(88)90100-9}{\emph{Nucl. Phys.}
  {\bfseries B310} (1988) 714}.

\bibitem{Sirlin:1985ux}
A.~Sirlin and R.~Zucchini, \emph{{Dependence of the Quartic Coupling H(m) on
  M($H$) and the Possible Onset of New Physics in the Higgs Sector of the
  Standard Model}},
  \href{https://doi.org/10.1016/0550-3213(86)90096-9}{\emph{Nucl. Phys.}
  {\bfseries B266} (1986) 389}.

\bibitem{Dib:2019iqo}
C.~Dib, S.~Kovalenko, I.~Schmidt and A.~Smetana, \emph{{Low-scale seesaw from
  neutrino condensation}}, \href{https://doi.org/10.1063/1.5130984}{\emph{AIP
  Conf. Proc.} {\bfseries 2165} (2019) 020023}.

\bibitem{Staub:2013tta}
F.~Staub, \emph{{SARAH 4 : A tool for (not only SUSY) model builders}},
  \href{https://doi.org/10.1016/j.cpc.2014.02.018}{\emph{Comput. Phys. Commun.}
  {\bfseries 185} (2014) 1773}
  [\href{https://arxiv.org/abs/1309.7223}{{\ttfamily 1309.7223}}].

\end{thebibliography}

\providecommand{\href}[2]{#2}\begingroup\raggedright\endgroup

\end{document}